\documentstyle[aps,epsf,multicol]{revtex}
\begin{document}
\draft
\title{Scale-free Network in Financial Correlations} 
\author{Hyun-Joo Kim$^1$, Youngki Lee$^2$, In-mook Kim$^1$ and Byungnam Kahng$^{3,*}$} 
\address{$^1$ Department of Physics, Korea University, Seoul 138-701, Korea \\
$^2$ Yanbian University of Science and Technology, Beishan St. Yanji, 
Jilin 133000, China \\
$^3$ School of Physics and Center for Theoretical Physics, Seoul National 
University, Seoul 151-742, Korea}
\date{\today}
\maketitle
\thispagestyle{empty}
\begin{abstract}
We study the cross-correlations in stock price changes between the S\&P 500 
companies by introducing a weighted random graph, where all 
vertices (companies) are fully connected, and each edge is weighted. 
The weight assigned to each edge is given by the normalized covariance 
of the two modified returns connected, so that it is ranged from -1 to 1. 
Here the modified return means the deviation of a return from its 
average over all companies. 
%The cross-correlation coefficients defined in this way turn out to be 
%time-independent. 
We define influence-strength at each vertex as the 
sum of the weights on the edges incident upon that vertex. 
Then we found that the influence-strength distribution in its absolute 
magnitude $|s|$ follows a power-law, $P(|s|)\sim |s|^{-\delta}$, with 
exponent $\delta \approx 1.8(1)$. 
\end{abstract}
\begin{multicols}{2}
\narrowtext
Recently complex systems such as biological, economic, physical, and 
social systems have received considerable attentions as an 
interdisciplinary subject\cite{science}. 
Such systems consist of many constituents such as individuals, 
companies, substrates, spins, etc, exhibiting cooperative and 
adaptive phenomena through diverse interactions between them. 
In particular, in economic systems, adaptive behaviors of 
individuals, companies, or nations, play a crucial role in forming 
macroscopic patterns such as commodity prices, stock prices, 
exchange rates, etc, which are formed mostly in a self-organized way 
\cite{arthur}. Recently, many attentions and studies have been 
focused and performed on the fluctuations and the correlations 
in stock price changes between different companies in physics 
communities by applying physics concepts and methods\cite{book1,book2}.\\ 

Stock price changes of individual companies are influenced by others. 
Thus, one of the most important quantities in understanding 
the cooperative behavior in stock market is the cross-correlation 
coefficients between different companies. Since the stock prices changes 
depend on various economic environments, it is very hard to construct 
a dynamic equation, and predict the evolution of the stock price change 
in the future. 
Recently, there have been some efforts to understand the 
correlations in stock price changes between different companies  
using a random matrix theory, where large eigenvalues are located 
far away from the rest part predicted by the random matrix theory, 
reflecting the collective behavior of the entire market as well as 
the subordination among the stock prices \cite{matrix1,matrix2,noh}.\\
   
Let $Y_i(t)$ be the stock-price of a company $i$ ($i=1,\dots, N$)
at time $t$. Then the return of the stock-price after a time 
interval $\Delta t$ is defined as  
\begin{equation}
S_i(t)=\ln Y_i(t+\Delta t)-\ln Y_i(t), 
\end{equation}
meaning the geometrical change of $Y_i(t)$ during the interval 
$\Delta t$. We take $\Delta t=1$ day for the following analysis 
in this Letter. The cross-correlations between individual 
stocks are considered in terms of a matrix $\bf C$, 
whose elements are given as 
\begin{equation}
C_{i,j}\equiv {{\langle S_i S_j \rangle - \langle S_i \rangle 
\langle S_j \rangle}\over 
{\sqrt{(\langle S_i^2\rangle-\langle S_i \rangle^2)
(\langle S_j^2 \rangle-\langle S_j \rangle^2)}}}, 
\end{equation}
where the brackets mean a temporal average over the period we studied. 
Then $C_{i,j}$ can vary between [-1,1], where $C_{i,j}=1$ (-1) means 
that two companies $i$ and $j$ are completely correlated 
(anti-correlated), 
while $C_{i,j}=0$ means that they are uncorrelated. Since the matrix 
$\bf C$ is symmetric and real, all eigenvalues are real, and the largest 
eigenvalue is not degenerate. It was found that the eigenvector 
corresponding to the largest eigenvalue is strongly localized at a 
few companies which strongly influence other companies in the 
stock price changes\cite{matrix1,matrix2}. 
Moreover, the ultrametric hierarchical tree structure was 
constructed among those companies, using the concept of the minimum 
spanning tree in the graph theory, which arranges those companies in 
order following their strengths of influence\cite{ultra,taxo}.\\

In this Letter, we study further properties of the cross-correlations 
in stock price changes using a random graph theory\cite{random}. 
The study of complex systems through a random graph was 
initiated by Erd\"os and R\'enyi (ER)\cite{er}, 
however, the ER model is too random to describe real-world complex 
systems. Recently, Barab\'asi and Albert (BA) \cite{ba} introduced 
an evolving network where the number of vertices increases 
with time rather than fixed, and a newly introduced vertex 
is connected to $m$ already existing vertices, following 
the so-called preferential attachment rule. 
The number of edges $k$ incident upon a 
vertex is called the degree of the vertex. Then the degree 
distribution $P(k)$ follows a power-law $P(k)\sim k^{-\gamma}$ 
with $\gamma=3$ for the BA model, while for the ER model, 
it follows a Poisson distribution. The network following a 
power-law in the degree distribution is called scale-free (SF) network.\\ 

For the problem of the correlations in stock price 
changes, each vertex (edge) in the random graph 
corresponds to a company (the cross-correlation in stock price 
changes between the companies connected via that edge).
The random graph generated in this way is different from a 
typical one in the following way: While the edge in a typical 
random graph has weight either 
1 or 0, depending on if the edge is present or not, respectively, 
the edge in the random graph we will introduce has weight 
$\{w_{i,j}\}$, which is rather distributed between [-1,1].
For further studies, the random graph for the former (later) case is 
called binary random graph (BRG) (weighted random graph (WRG)).
While WRG can be found easily in real-world networks such 
as neural networks, cardiovascular networks, and respiratory networks 
in biological systems, acquaintance networks in social systems, etc, 
it is less studied, compared with BRG\cite{review1,review2}. 
In WRG, one may wonder if the edge with weak weight, called the 
weak edge, can be ignored in considering the correlations, 
however, there have been ongoing discussions about the importance 
of the weak edges, for example, in social acquaintance network, the 
scientific collaboration web, and ecosystems\cite{ecosystem,yook}. 
We will also show that the WRG introduced in this Letter is 
different from the corresponding DRG in which weak edges 
are ignored and the remaining edges are assigned a unit weight. \\ 

Recently, Yook, Jeong and Barab\'asi (YJB) introduced a WRG\cite{yook}. 
In this model, a vertex $i$ is newly introduced at each time step, 
connecting to $m$ vertices existing already according to the 
so-called preferential attachment rule. The edge connecting from 
the vertex $i$ to an existing vertex $j$ is assigned a weight 
$w_{i,j}$, depending on the degree of the vertex $j$, which is 
not uniform. The weight at each vertex is defined as the sum 
of the weights on the edges incident upon that vertex, which 
follows a power-law in its distribution, $P(w)\sim w^{-\delta}$, 
where $w$ means weight at a vertex. 
The exponent $\delta$ is different from the degree exponent 
$\gamma$, and turns out to strongly depend on the mean degree $m$. 
While the YJB model is meaningful as 
a first step towards understanding diverse WRGs in real-world 
in a simple way, 
it is very different from what we will consider in association 
with the cross-correlations in stock price changes. 
The WRG we will introduce below are fully connected, and 
each edge is weighted, while the YJB graph is also weighted, but 
sparsely connected.\\

\begin{figure}
\epsfxsize=8cm
\epsfysize=8cm
\epsffile{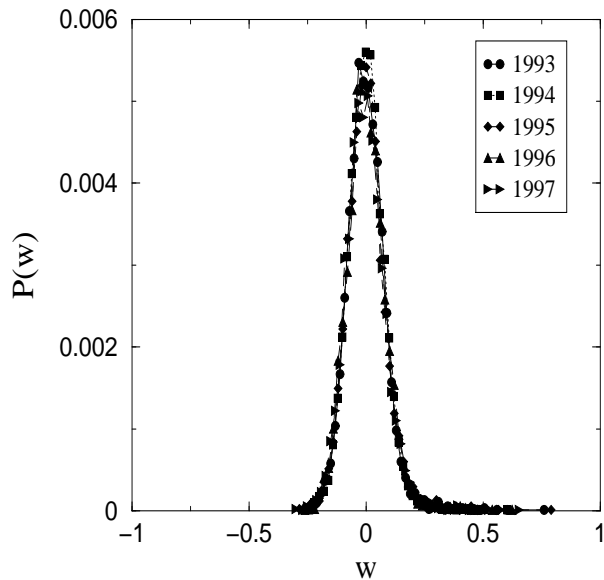}
\caption
{Plot of the distribution of the cross-correlation coefficients 
$\{w_{i,j}\}$. The data are obtained by temporal averaging over 
each year from 1993 to 1997.}   
\label{rd}
\end{figure}

We consider the cross-correlations in stock-price changes 
between the S\&P 500 companies during 5-year period 1993-1997.
Thus, the $N=500$ companies correspond to 500 vertices, which 
are fully connected to each other through $N(N-1)/2$ edges. 
Each edge is assigned a weight, $w_{i,j}$ ($i,j=1,\dots,N$), 
which is slightly modified from the cross-correlation coefficient 
$C_{i,j}$ defined in Eq.(2). Before defining $w_{i,j}$ specifically, 
we first recall some properties exhibited by $C_{i,j}$. It is known that 
the distribution of the coefficients $\{C_{i,j}\}$ is a bell-shaped 
curve, and the mean value of the distribution is slowly 
time-dependent, while the standard deviation is almost constant
\cite{conf}. 
The time-dependence of the mean value might be caused by external 
economic environments such as bank interest, inflation index, 
exchange rate, etc, which fluctuates from time to time. 
To extract intrinsic properties of the correlations in stock price 
changes, we introduce a quantity,    
\begin{equation}
G_i(t)=S_i(t)-{1\over N}\sum_i S_i(t), 
\end{equation}  
where $G_i(t)$ indicates the relative return of a company $i$ to its 
mean value over the entire 500 companies at time $t$. 
The cross-correlation coefficients are redefined in terms of $G_i$, 
\begin{equation}
w_{i,j}\equiv {{\langle G_i G_j \rangle - \langle G_i \rangle 
\langle G_j \rangle}\over 
{\sqrt{(\langle G_i^2\rangle-\langle G_i \rangle^2)
(\langle G_j^2\rangle-\langle G_j \rangle^2)}}}. 
\end{equation}
The cross-correlation coefficients $\{w_{i,j}\}$ are assigned 
to each edge of the WRG as its weight. 
In order to check if the distribution $P(w)$ 
is time-independent, we take temporal average in Eq.(4) over each 
year from 1993 to 1997. In Fig.\ref{rd}, 
we plot the distributions of $\{w_{i,j}\}$ obtained for each year. 
The data for different years are indeed overlapped, and 
time-independent. Therefore we think that 
the cross-correlation coefficients $\{w_{i,j}\}$ we introduced 
are appropriate to study intrinsic properties of the cross-correlations 
between the 500 S\&P companies. \\ 

\begin{figure}
\epsfxsize=8cm
\epsfysize=8cm
\epsffile{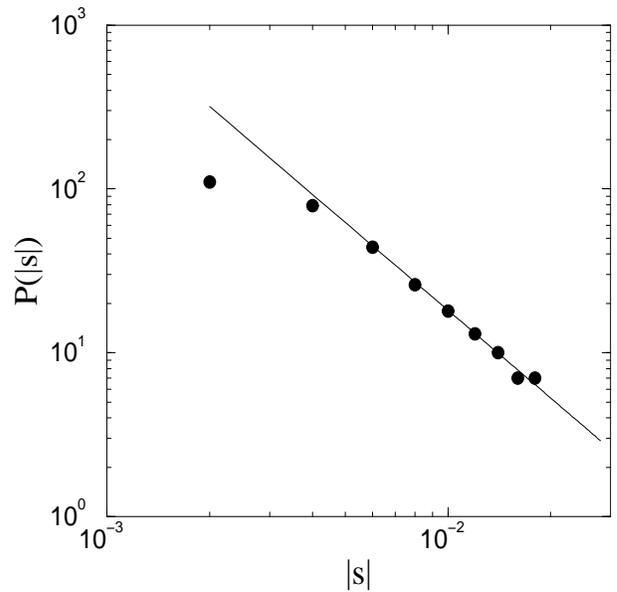}
\caption
{Plot of the influence-strength distribution $P(|s|)$ versus 
the absolute magnitude of the influence-strength $|s|$ in 
double-logarithmic scales. The solid guideline has 
a slope $-1.8$.}
\label{wd}
\end{figure}

We define the influence-strength $s_i$ at a vertex $i$ as the sum 
of the weights on the edges incident upon the vertex $i$, that is, 
\begin{equation}
s_i=\sum_{j\ne i} w_{i,j},
\label{tw}
\end{equation} 
where $j$ denotes the vertices connected to the vertex $i$. 
Here $\{w_{i,j}\}$ was obtained numerically by temporal averaging 
over the 5 years in Eq.(4). Then the weight $s_i$ means the net amount 
of influence for the company $i$ to affect other companies in stock 
price changes. Since the weight $w_{i,j}$ is distributed between 
[-1,1], the influence-strength at a certain vertex could be negative. 
Thus, we deal with the absolute magnitude of the influence-strength 
for each vertex. 
In Fig.\ref{wd}, we plot the influence-strength distribution 
$P(|s|)$ in the absolute magnitude as a function of $|s|$, which 
turns out to follow a power-law, $P(|s|) \sim |s|^{-\delta}$. 
The exponent $\delta$ is estimated to be $\delta \approx 1.8(1)$. 
Thus the cross-correlations in stock price changes forms a SF 
network, which in particular is a weighted SF network.
To our knowledge, this is the first observation of SF 
WRG emerging in real-world economic systems.
The presence of a scaling in the cross-correlations through 
the WRG would be related to the ultrametric hierarchical tree structure, 
implying that there exist a few companies having strong influence in stock 
price changes. 
On the other hand, it is easy to know that as the degree exponent 
in SF networks is smaller, the connectivity to the hub, the vertex 
with the largest degree, is higher, and the network is much centralized. 
This fact is also applicable to the SF WRG we introduced. 
Since the influence-strength 
exponent $\delta$ is smaller than 2 in the WRG, the vertex having 
largest influence-strength plays a much important role in affecting 
stock price changes of other vertices, compared with the role of 
the hub in the Internet\cite{internet}, the world-wide web\cite{www} and 
the metabolic network\cite{metabolic}, where the degree exponent 
is greater than 2. We think that this result reflects that 
economic systems is much correlated and adaptive to achieve high 
profits. Thus the rich-get-richer phenomenon appears much strongly 
in economic systems, compared with the information systems.  
In contrast, we can expect that a simple drop in the stock price 
occurring in one of most influential companies could lead to a 
crash in the entire stock market.\\ 

\begin{figure}
\epsfxsize=8cm
\epsfysize=8cm
\epsffile{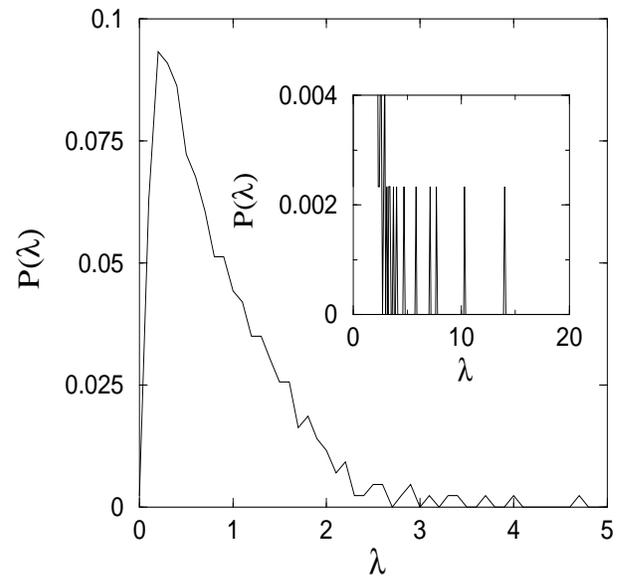}
\caption
{Plot of the distribution of the eigenvalues of the matrix $\bf W$.
Inset: Plot of the distribution of the eigenvalues versus eigenvalue 
in the fat-tail region.}
\label{ed}
\end{figure}

Next, we study the matrix $\bf W$ whose elements are given by 
$w_{i,j}$. We found the distribution of the eigenvalue 
$\lambda$ for the matrix $\bf W$ shown in Fig.\ref{ed}, looking 
similar to the one for the matrix $\bf C$ previously 
studied\cite{matrix1,matrix2}; 
For small $\lambda$, the spectrum is likely to follow the 
prediction by the random matrix theory, while it exhibits 
a fat-tail behavior for large $\lambda$. However, we found 
that the gap between the first and second largest eigenvalues 
for the matrix $\bf W$ is not as big as the one for the matrix 
$\bf C$. Thus we think that the huge gap observed in the 
eigenspectrum for the matrix $\bf C$ may be caused by some 
external effects. \\ 

Finally, we also consider the components $\{v_{j,1}\}$ of the 
eigenvector corresponding to the largest eigenvalue for the 
matrix $\bf W$, where the vertices are ordered following the 
absolute magnitude of influence-strength. According to the matrix theory, 
the square of the component, $\{v^2_{j,1}\}$, indicates the 
relative strength of the contributions for each vertex to the 
largest eigenvalue. We plot the square of the component of 
the eigenvector $\{v^{2}_{j,1}\}$ versus the index $j$ 
in Fig.4. It is found that the components $\{v^{2}_{j,1}\}$ 
is strongly localized at the vertices with strong influence-strength 
as shown in Fig.4. Thus, the companies with large influence-strength 
affect other companies strongly in stock price changes. 
Interestingly, we found another SF behavior, 
$v^2_{j,1}\sim j^{-\beta}$ with $\beta \approx 1.8(2)$ for large $j$, 
as shown in the inset of Fig.4, which is consistent with the one 
$\delta\approx 1.8(1)$, obtained from the influence-strength distribution.\\ 

It would be interesting to compare our result with the one 
by Vandewalle et al.\cite{mst}, who eliminated 
weak edges, and regarded the remaining sparse random graph as 
a BRG. To be specific, they considered the cross-correlation 
coefficients defined in Eq.(2) in stock price changes between 
the 6 358 US companies, and assigned them to each edge of the fully 
connected graph. Next, they chose $N$ strongest edges out of 
$N(N-1)/2$ edges in the magnitude of $C_{i,j}$, 
and assigned a unit strength to each of them, so that weak edges 
are ignored. Then the graph becomes sparse and binary. 
They measured the degree distribution for this BRG, which follows 
a power-law, $P(k)\sim k^{-\gamma}$, with $\gamma\approx 2.2$, 
which is obviously different from our results, $\delta\approx 1.8(1)$ and 
$\beta \approx 1.8(2)$.\\ 

In conclusions, we have considered the cross-correlations in stock 
price changes between the S\&P 500 companies by introducing a 
weighted random graph (WRG). The vertices of the WRG representing 
the 500 companies are fully connected to each other through 
weighted edges. An edge connecting vertices 
$i$ and $j$ has the weight given by the correlation coefficient $w_{i,j}$, 
defined as the normalized covariance of modified returns of 
two companies $i$ and $j$. Here the modified return of a company 
$i$ means the deviation of the return of the company $i$ from its 
average over all 500 companies. This modification yields 
the effect of excluding the overall behavior of the entire stock 
prices fluctuating from time to time. 
The distribution of the correlation coefficients obtained using 
the modified return is time-independent, and the coefficients 
themselves describe generic correlations between different 
companies without considering the effect of external environments. 
Next, we defined the influence-strength at each 
vertex as the sum of the weights assigned to the edges incident 
upon that vertex. We found that the influence-strength distribution 
follows a power-law $P(|s|)\sim |s|^{-\delta}$ 
with $\delta \approx 1.8(1)$, where $s$ means influence-strength. 
The fact that the exponent $\delta$ is smaller than 2 
implies that the stock price changes of the 500 companies are much 
strongly correlated, compared with the Internet topology, or 
the world-wide web, reflecting that cooperative and adaptive phenomena 
appear much dominantly in economic systems.  
%We also studied the eigenspectrum of the matrix consisting of 
%the cross-correlation coefficients $\{w_{i,j}\}$. It was found 
%that the eigenvector corresponding to the largest eigenvalue is 
%strongly localized at the vertice with largest influence-strength.
%Moreover, the square of the eigenvector component $v_{j,1}^2$ 
%($j=1\dots N$) behaves as $v^2_{j,1}\sim j^{-\beta}$ 
%with $\beta\approx 1.8(2)$, when the vertices 
%are ordered following the absolute magnitude of its 
%influence-strength. \\ 

This work is supported by Korean Research Foundation grant 
(KRF-2001-015-DP0120) and by the Ministry of Education 
through the BK21 project in KU, and by grants 
No.2000-2-11200-002-3 from the BRP program of the KOSEF 
through SNU. 

%%%%%%%%%%%%%%%%%%%%%%%%%%%%%%%%%%%%%%%%%%%%%%%%%%%%%%%%%%%%%%%%%%%%%%%%%%%%

%%%%%%%%%%%%%%%%%%%%%%%%%%%%%%%%%%%%%%%%%%%%%%%%%%%%%%%%%%%%%%%%%%%%%%%%%%%%
%%%%%%%%%%%%%%%%%%%%%%%%%%%%%%%%%%%%%%%%%%%%%%%%%%%%%%%%%%%%%%%%%%%%%%%%%%%%
%
%  Figures
%
%%%%%%%%%%%%%%%%%%%%%%%%%%%%%%%%%%%%%%%%%%%%%%%%%%%%%%%%%%%%%%%%%%%%%%%%%%%%
%%%%%%%%%%%%%%%%%%%%%%%%%%%%%%%%%%%%%%%%%%%%%%%%%%%%%%%%%%%%%%%%%%%%%%%%%%%%

%%%%%%%%%%%%%%%%%%%%%%%%%%%%%%%%%%%%%%%%%%%%%%%%%%%%%%%%%%%%%%%%%%%%%%%%%%%%
%%%%%%%%%%%%%%%%%%%%%%%%%%%%%%%%%%%%%%%%%%%%%%%%%%%%%%%%%%%%%%%%%%%%%%%%%%%%
\begin{figure}
\epsfxsize=8cm
\epsfysize=8cm
\epsffile{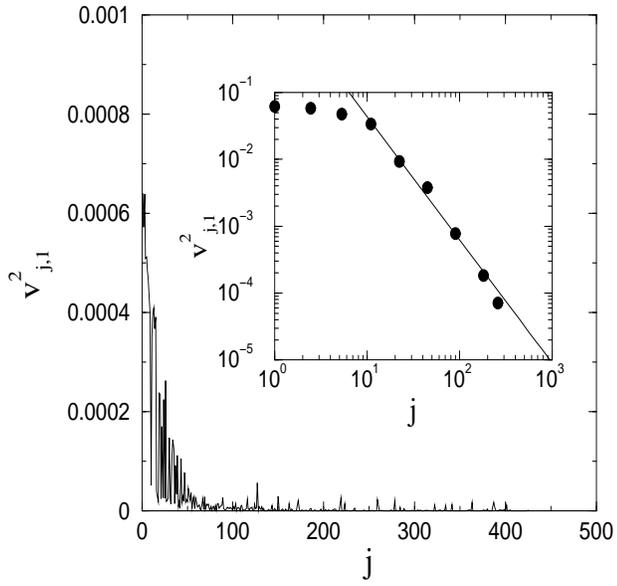}
\caption
{Plot of the component squares $v_{j,1}^2$ of the eigenfunction 
corresponding to the largest eigenvalue versus the vertex index 
$j$. The vertices are ordered following the absolute magnitudes 
of their influence-strengths. Inset: The same plot as the main panel 
but in double-logarithmic scales using log-bin. 
The straigth guideline represents $v^{2}_{j,1} \sim j^{-1.8}$.}
\label{vec}
\end{figure}  
%%%%%%%%%%%%%%%%%%%%%%%%%%%%%%%%%%%%%%%%%%%%%%%%%%%%%%%%%%%%%%%%%%%%%%%%%%%%
\end{multicols}
\end{document}